\documentclass[aps,pra,reprint,superscriptaddress]{revtex4-2}
\usepackage[T1]{fontenc}
\usepackage[utf8]{inputenc}

\usepackage{graphicx}% Include figure files
\usepackage{dcolumn}% Align table columns on decimal point
\usepackage{bm}% bold math
\usepackage{hyperref}% add hypertext capabilities
\usepackage{inputenc}
\usepackage{calc}
\usepackage{amsmath, amssymb, amsthm}
\usepackage{xcolor,multirow, tikz}
\usepackage{hyperref, cleveref, url}
\usepackage{amsfonts}
\usepackage{bbold}
\usepackage{blkarray}
\usepackage{latexsym}
\usepackage{mathrsfs}
\usepackage{thmtools}
\usepackage{mathtools}
\usepackage{pgfplots}
\usepackage{physics}
\usepackage{comment}

\usepackage{cancel}
\usepackage{ulem}
\usepackage{tcolorbox} % colored box! 
\usepackage{float}

\newcommand{\br}{{\bf r}}

\newcommand{\bd}{{\bf d}}
\newcommand{\E}{{\bf E}}

\newcommand{\om}{\omega}
\newcommand{\Om}{\Omega}
\newcommand{\be}{\begin{equation}}
\newcommand{\ee}{\end{equation}}
\NewDocumentCommand{\scinot}{m m o}{%
    \ensuremath{%
        #1 \times 10^{#2}%
        \IfValueT{#3}{\,\mathrm{#3}}%
    }%
}

\begin{document}

\title{Dynamical Casimir photons from rotation of a nonspherical particle}

\author{Guilherme C. Matos}
 \email{gcmatos@ufscar.br}
 \affiliation{Instituto de F\'{\i}sica, Universidade Federal do Rio de Janeiro \\ Caixa Postal 68528,   Rio de Janeiro,  Rio de Janeiro, 21941-972, Brazil}
 \affiliation{Departamento de F\'{\i}sica, Universidade Federal de S\~ao Carlos \\ S\~ao Carlos, S\~ao Paulo, 13565-905, Brazil}
 \author{Lucas Bianchi}
 \email{lucasbnchm@gmail.com}
\affiliation{Instituto de F\'{\i}sica, Universidade Federal do Rio de Janeiro \\ Caixa Postal 68528,   Rio de Janeiro,  Rio de Janeiro, 21941-972, Brazil}
\author{Jeremy N. Munday}
%\email{jnmunday@ucdavis.edu}
\affiliation{Department of Electrical and Computer Engineering, University of California, Davis, CA 95616, USA}
\author{François Impens}
%\email{impens@if.ufrj.br}
\affiliation{Instituto de F\'{\i}sica, Universidade Federal do Rio de Janeiro \\ Caixa Postal 68528,   Rio de Janeiro,  Rio de Janeiro, 21941-972, Brazil}
\author{Reinaldo de Melo e Souza}
%\email{reinaldos@id.uff.br}
\affiliation{Instituto de Física, Universidade Federal Fluminense, 24210-346, RJ, Brazil}
\author{Paulo A. Maia Neto}
 \email{pamn@if.ufrj.br}
\affiliation{Instituto de F\'{\i}sica, Universidade Federal do Rio de Janeiro \\ Caixa Postal 68528,   Rio de Janeiro,  Rio de Janeiro, 21941-972, Brazil}%

\date{\today}% 

\begin{abstract}
We consider a non-spherical neutral particle spinning in free space and interacting with the electromagnetic quantum vacuum. When the rotation axis is orthogonal to the particle symmetry axis, the scattered field develops frequency sidebands that induce the parametric emission of dynamical Casimir photon pairs. Under the structural constraint of a maximum tip velocity, the emission rate is maximized for a nearly spherical geometry and is further enhanced near a polaritonic resonance. For realistic material parameters, even these optimized upper bounds remain exceedingly small, setting stringent quantitative limits on free-space rotational dynamical Casimir emission with a single nanoparticle.
\end{abstract}

\maketitle

\section{Introduction}

Recent experiments with optically levitated dielectric nanoparticles have achieved rotation frequencies beyond the GHz range~\cite{Ahn2018,Reimann2018,Ahn2020,Jin2021,Zielinska2024,Peng2023}, opening a new regime of ultrafast mechanical motion. Such frequencies may enable the observation of vacuum effects associated with rotation, including rotational quantum friction near surfaces~\cite{Ju2023,Peng2023,Zhao2012,Xu2021}. More broadly, the interplay between rotation and quantum vacuum physics has become a topic of growing interest, giving rise to phenomena such as quantum friction~\cite{Manjavacas10,Maghrebi2012}, enhanced heat transfer~\cite{Sanders2019}, a quantum-vacuum analogue of the Sagnac effect~\cite{Matos2021}, and nonequilibrium Casimir forces~\cite{Amaral2025}. In this work, we investigate a distinct mechanism: the dynamical Casimir emission (DCE) of photon pairs generated by a spinning anisotropic particle rather than by an oscillating boundary~\cite{Lambrecht1996,Jaekel1997,Kardar1999,Maghrebi2013,Farias2019, Dodonov2020,Gong2021,Woods2021,Impens2022}.

The recent progress in ultrafast levitated rotation naturally raises the question of whether DCE could be observed in such experimental platforms. So far, the small magnitude of oscillation-induced DCE under realistic conditions has hindered its direct observation, although an analog realization has been demonstrated in circuit QED~\cite{Wilson2011}.

In the present setting, dynamical Casimir emission originates from the frequency modulation of the anisotropic scattering response induced by rotation. When the rotation axis is not a symmetry axis, sidebands couple positive- and negative-frequency field components, leading to a Bogoliubov mixing between annihilation and creation operators~\cite{Dodonov1990,Dodonov1992,MaiaNeto1996,Lombardo25} and hence to photon creation from the vacuum. The spectral range involved in this spinning-induced DCE increases with the rotation frequency $\Omega$, resulting in a strong dependence of the emission rate on the spinning frequency. In contrast to quantum friction~\cite{Manjavacas2010,Maghrebi2012,Milton2016,Reiche2022,Oue2025,Pereira2026}, the mechanism considered here does not require a lossy material medium.

\begin{figure}[htbp]
    \centering
    \includegraphics[width=6 cm]{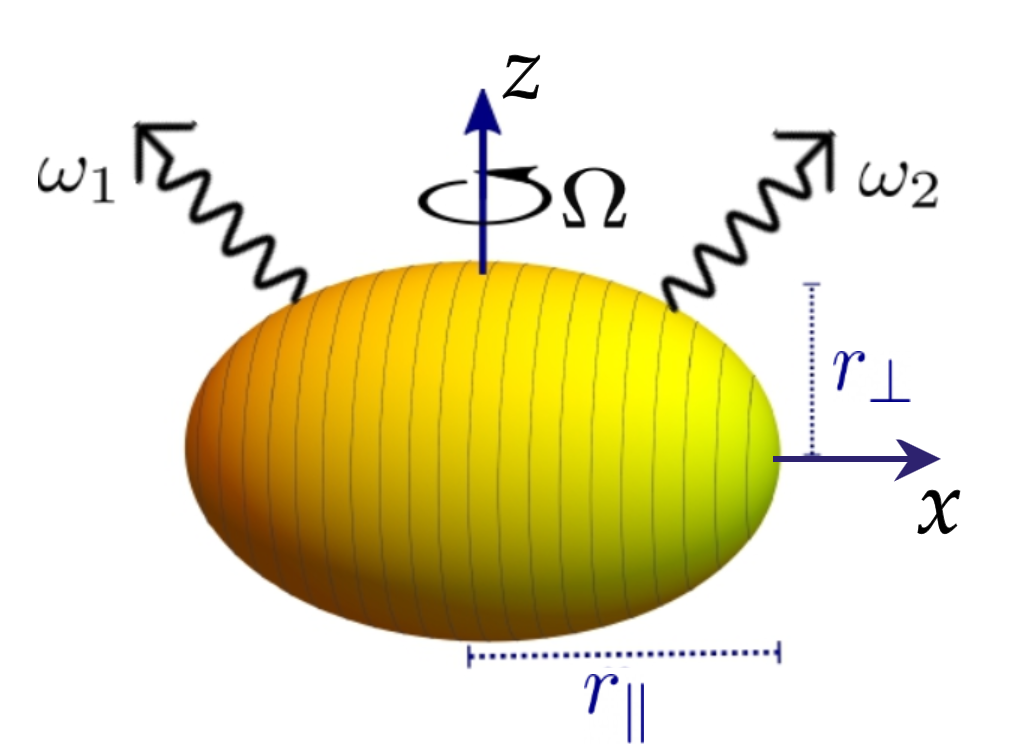} %[width=0.8\linewidth]{Images/esferoide 3d.png}
    \caption{Dynamical Casimir Effect (DCE) radiation from a rotating spheroid. The spheroid rotates at angular velocity $\Omega$ about the $z$-axis, which is orthogonal to its symmetry axis (along the $x$ direction). The particle's anisotropic polarizability enables rotational sideband mixing of vacuum fluctuations, driving the emission of  photon pairs that satisfy the  condition $\omega_1+\omega_2=2\Omega$.} 
    \label{spheroid}
\end{figure}

Preliminary studies of spinning-induced DCE were carried out using a two-dimensional scalar-field toy model~\cite{Maghrebi2013}. However, assessing the feasibility of such experiments requires a complete theory in a realistic three-dimensional setting. Here we develop such a framework, including a full treatment of the quantized electromagnetic field together with a realistic dispersive material response. We show how the emission can be enhanced by tuning the rotation frequency close to the GHz polaritonic resonance of a Barium Strontium Titanate (BST) nanoparticle~\cite{Xu2021,Turky2015,Turky2016}. As the simplest anisotropic geometry, we consider a spheroidal particle rotating about an axis orthogonal to its symmetry axis. We find that the eccentricity maximizing the DCE emission has a nontrivial dependence on particle size.  Because the attainable rotation frequency is fundamentally limited by the maximum tip velocity supported by the material, we also derive geometry-optimized upper bounds for the DCE emission rate for a given material.

 In Sec.~II, we derive the emission spectrum and total photon-production rate in terms of the polarizability tensor of the rotating particle. In Sec.~III, we apply the theory to BST spheroids and identify the geometric and resonant conditions that maximize the effect under realistic constraints. Section~IV concludes with a discussion of the observability limits of free-space rotational dynamical Casimir emission.

\section{DCE frequency spectrum for a spinning particle}

 Here we develop the general theoretical framework for dynamical Casimir emission from a spinning nano-object and show that it depends critically on the particle anisotropy. 
  We derive the scaling of DCE emission with the rotation frequency and compare it with known DCE results: $\Omega^5$ for an oscillating plane mirror of large transverse size~\cite{MaiaNeto1996,Ford1982,MaiaNeto1994}, and $\Omega^9$ for subwavelength mirrors~\cite{Alonso2024}, oscillating spherical particles~\cite{MaiaNeto1993}, and oscillating ground-state atoms~\cite{MeloeSouza2018}.

\subsection{Input-output transformation in the dipolar regime}

 Within the dipole approximation, the scattering by a rotating nanoparticle (located at the origin) yields the following input-output relation for the electric field~\cite{MaiaNeto1996}:
\be
    \vb{E}_{\rm out}(\br,\om) = \mathbf{G}(\br,\br'=0,\om)\cdot\bd(\om)+\vb{E}_{\rm in}(\br,\om),
    \label{inout}
\ee
where $\bd(\om)$ is the particle's induced electric dipole while $\mathbf{G}= \mathbf{G}^R- \mathbf{G}^A$ is the difference between the retarded and advanced Green's functions. In the radiation zone, the latter is given by
\be
    \mathbf{G}(\br,\br'=0,\om) = \frac{2i\om^2\sin\left(\frac{\om r}{c}\right)}{4\pi \epsilon_0 c^2 r}\left[\mathbb{I}-\frac{\br\otimes\br}{r^2}\right].\label{asymG}
\ee

 For a localized scatterer, it is convenient to expand the input and output fields as a sum over multipolar spherical waves~\cite{Berestetski1972,cohen1992}:
\begin{eqnarray}
    \vb E_{\rm in}(\vb r,\om)&=&i\mathcal{E}(\om)\sum_{jm\lambda}\Big[a_{jm\lambda}^{\rm in} (\om)\Theta(\om)\boldsymbol{I}_{jm\lambda}(\vb r, \om)\nonumber\\
    & &-a_{jm\lambda}^{\rm in \dagger}(-\om)\Theta(-\om)\boldsymbol{I}^*_{jm\lambda}(\vb r, \om)\big],\label{esph}
\end{eqnarray}
and likewise for the output field $\vb E_{\rm out}(\vb r,\om)$, where 
$\Theta$ denotes the Heaviside step function and $\mathcal{E}(\om) = \sqrt{\frac{\hbar \om}{2\epsilon_0c}}.$  Here, $j=1,2,\ldots$ denotes the multipole order (equivalently, the total angular-momentum quantum number), while $m=-j,\ldots,j$ corresponds to its projection along the $z$ axis. The spherical-wave modes $\boldsymbol{I}_{jm\lambda}(\vb r,\om)$~\cite{cohen1992} are labeled by $\lambda=\mathrm{E},\mathrm{M}$, corresponding to electric and magnetic multipoles respectively. In the radiation zone, the electric multipole modes can be approximated by:
\begin{equation}\label{asym}
    \boldsymbol{I}_{jm{\rm E}}(\vb r, \om)\approx\frac{4\pi i^{j-1}}{\sqrt{j(j+1)}}\frac{\cos{\qty (\frac{\om r}{c}-\frac{j\pi }{2})}}{r}\boldsymbol{\nabla}Y_{j m}(\hat{\vb r}),
\end{equation}
where $Y_{jm}(\hat{\vb r})$ denote the spherical harmonics.  The operators $a_{jm\lambda}^{\rm in}(\om)$ and $a_{jm\lambda}^{\rm in\dagger}(\om)$ of the quantized light field obey the standard commutation relations.

We derive a Bogoliubov transformation between input and output bosonic operators, mediated by the dipole moment of the spinning particle, by replacing the expansion (\ref{esph}) into
(\ref{inout}) and taking the asymptotic expressions 
(\ref{asymG}) and
(\ref{asym})
into account. As expected from selection rules, only the electric dipole modes ($j=1$; $m=0,\pm1$, E) are excited:
    \begin{eqnarray}
         a^{\rm out}_{10{\rm E}}(\om)& = &\frac{\om^2}{2\sqrt{6}\pi^{3/2}\epsilon_0 c^2\mathcal{E}(\om)}d_z(\om)+a^{\rm in}_{10{\rm E}}(\om)\nonumber\\
         a^{\rm out}_{1 \pm1 {\rm E}}(\om) & = & \frac{\mp\om^2}{4\sqrt{3}\pi^{3/2}\epsilon_0 c^2\mathcal{E}(\om)}(d_x(\om)\mp id_y(\om))\nonumber \\
        & & + \,a^{\rm in}_{1\pm1{\rm E}}(\om)\label{bogo}
    \end{eqnarray}

\subsection{Spectrum and total emission rate} \label{scattring-section}

 The number of photons emitted into mode $(j=1,m,\mathrm{E})$ within the frequency interval $[\omega,\omega+d\omega]$, per interaction time $T$, is
%The number of photons emitted in  mode $j,m,{\rm E}$ in the frequency
%interval $[\omega,\omega+d\omega]$ divided by the interaction time $T$ is given by
\begin{equation}\label{Ndef}
     d\Gamma_{m}=\frac1T\bra{0_{\rm in}}a^{\rm out\dagger}_{1m{\rm E}}(\om) a^{\rm out}_{1m{\rm E}}(\om)\ket{0_{\rm in}}\,d\omega.
\end{equation}
 The corresponding spectra of emission are given by
    \begin{eqnarray} \label{N}
        \frac{d\Gamma_{0}}{d\om} &=& \frac{1}{12\pi^3}\frac{\om^3}{\hbar c^3\epsilon_0}C_{zz}(\omega)\\
\label{spec11}
        \frac{d\Gamma_{\pm1}}{d\om}&=& \frac{1}{48\pi^3}\frac{\om^3}{\hbar c^3\epsilon_0}[C_{xx}(\omega)+C_{yy}(\omega)\\
        \nonumber
        & &\pm i(C_{yx}(\omega)-C_{xy}(\omega))],
    \end{eqnarray}
where 
\be  
    \label{spectrumfluc}C_{\mu \nu}(\omega) = \bra{0_{\rm in}}d_{\mu}^\dagger(\om)d_{\nu}(\om)\ket{0_{\rm in}}
\ee
are the elements of the
dipole correlation matrix ($\mu,\nu$ denoting Cartesian components). 

We now consider a nanoparticle rotating with angular frequency $\Omega$ about the $z$ axis, orthogonal to the symmetry axis taken along $x$~(Fig.~\ref{spheroid}). The induced dipole of the particle at rest is $\bd_{\rm rest}(\om) =\boldsymbol{\alpha}_{\rm rest}(\om)\cdot\E(0,\om)$, with the polarizability tensor
\be 
   \boldsymbol{\alpha}_{\rm rest}(\om) = \begin{pmatrix}
    \alpha_{\parallel}(\om) & 0 & 0\\
0 & \alpha_\perp(\om) & 0\\
0 & 0 & \alpha_{\perp}(\om)
    \end{pmatrix}.
    \label{alpharest}
\ee 
Rotation modulates the polarizability and generates sidebands at frequencies $\omega\pm2\Omega$, because a full modulation cycle corresponds to a half turn of the particle. Under these conditions, 
the induced electric dipole is given by 
(see Appendix A) 
\be 
    \begin{split}
        \bd(\om)  =  &\,\boldsymbol{\alpha}_0(\omega)\cdot\vb{E}_{\rm in}(\mathbf{0},\omega)\\
        &+\sum_{\sigma=\pm}\boldsymbol{\alpha}_{\sigma}(\omega)\cdot\vb{E}_{\rm in}(\mathbf{0},\omega+2\sigma \Omega).
    \end{split}
    \label{drot}
\ee  %\ref{alphapm},\ref{Deltadef
The dipole correlation matrix $C_{\mu\nu}(\omega)$ is obtained from Eqs.~\eqref{spectrumfluc} and \eqref{drot}. Owing to the vacuum expectation value~\eqref{spectrumfluc}, it involves only correlations between red- and blue-detuned sideband components of the dipole, associated with frequency shifts $\omega\mp 2\Omega$. Thus, the polarizability tensor $\boldsymbol{\alpha}_0(\omega)$ capturing the dipole response at the electric field frequency is irrelevant for DCE. By contrast, the sideband generation and DCE emission depends crucially on the Doppler-shifted susceptibility tensors
\begin{eqnarray}
        \boldsymbol{\alpha}_{\pm}(\omega) &=&\frac{\Delta(\omega\pm \Omega)}{2} \begin{pmatrix}
        1 & \mp i & 0\\
        \mp i&  -1 & 0\\
        0 & 0 & 0
        \end{pmatrix}\label{alphapm}
\end{eqnarray}
 which scale with the geometry-dependent anisotropic response 
\begin{equation}\label{Deltadef}
\Delta(\omega) =  \frac{\alpha_\parallel(\omega)-\alpha_\perp(\omega)}{2}.
\end{equation}
In particular, $\boldsymbol{\alpha}_{\pm}(\omega) $ vanish if the particle response is isotropic in the plane perpendicular to the rotation axis,
suppressing Bogoliubov mixing in this case. While a rotating isotropic nanoparticle may exhibit other rotation-induced effects~\cite{Manjavacas10,Manjavacas2010,Matos2021,Amaral2025}, anisotropy in the plane orthogonal to the rotation axis is a necessary condition for spinning-induced DCE.

As expected, $d_z(\omega)$ is unaffected by rotation about the $z$ axis. This reflects a selection rule, since  $\boldsymbol{\alpha}_\pm$ couple only the $x$ and $y$ components [Eq.~\eqref{alphapm}] -- modes with $m=0$ do not contribute to the emission. In addition, radiation in the $m=-1$ channel vanishes in Eq.~\eqref{spec11} by destructive quantum interferences. Hence, DCE radiation occurs only into the mode $m=1$, so that $d\Gamma \equiv d \Gamma_{1}$. The emission spectrum is given by:
\be\label{spectrumEq}
    \frac{d\Gamma}{d\om} = \frac{|\om-2\Omega|^3\om^3|\Delta(\om-\Omega)|^2\Theta(2\Omega-\om)}{36\pi^3c^6\epsilon^2_0}.
\ee
 The DCE spectrum is confined to the range $[0,2\Omega]$, where rotational modulation generates a negative-frequency sideband that enables Bogoliubov mixing.
Furthermore, the invariance of Eq.~\eqref{spectrumEq} under $\omega\to2\Omega-\omega$  shows that the spectrum is mirror-symmetric about the midpoint $\omega=\Omega$. This is indeed a fundamental property of DCE, which emits photons in pairs with frequencies satisfying $\omega_1+\omega_2=2\Omega$~\cite{Lambrecht1996,MaiaNeto1996,MeloeSouza2018}.
%.

The total emission rate is obtained by integrating the emission spectrum \eqref{spectrumEq} over all frequencies:
\begin{equation}\label{Gamma}
    \Gamma(\Omega)=\frac{1}{144\pi^3 c^6\epsilon^2_0}\int_0^{2\Omega}d\om \om^3 |\om-2\Omega|^3|\Delta(\om-\Omega)|^2.
\end{equation}
Equations~\eqref{spectrumEq} and \eqref{Gamma} constitute the central results of this work. They show that the core ingredient for rotational DCE is the anisotropy function $\Delta(\omega)$, which encodes the shape-induced difference between the parallel and transverse polarizabilities.
Those polarizabilities are determined by the
the dielectric function $\epsilon(\omega)$ of the particle's material.
For most dielectric materials, 
the resonant frequencies are all much higher than the rotation frequency $\Omega,$ allowing us to neglect dispersion and approximate $\Delta(\omega)$
by its zero-frequency value over the spectral range limited by $2\Omega$ in \eqref{Gamma}. The emission rate can then be approximated by its quasi-static limit
\begin{equation}
    \Gamma_{\rm qs}=\frac{2}{315\pi^3c^6\epsilon^2_0}\,|\Delta(0)|^2\,\Om^7.
    \label{gammaqs}
\end{equation}

 As shown below, for the GHz rotation frequencies achieved with optically levitated SiO$_2$ nanodumbbells~\cite{Ahn2018,Reimann2018,Ahn2020,Jin2021}, the quasi-static regime applies and yields an extremely weak DCE emission.
To circumvent this limitation, we investigate materials with polaritonic resonances in the GHz range, which enable to overcome the quasi-static regime and provide an enhanced emission rate.

 \section{DCE radiation from a spinning spheroid}
%prolate
As a simple example of an anisotropic object, we consider a prolate spheroid rotating about an axis perpendicular to its symmetry axis (Fig.~\ref{spheroid}) and obtain the corresponding DCE emission rate
 using the general theory developed in Sec.~II.  We first analyze the anisotropy function for this geometry and briefly discuss the case of SiO$_2$ nanoparticles, then investigate resonant enhancement in a BST material, and finally optimize the nanoparticle geometry to maximize the emission at fixed tip velocity.

%\begin{figure}[htbp]
%    \centering
%    \includegraphics[width=6 cm]{Images/DCE_GraphicalAbstract.pdf} %[width=0.8\linewidth]{Images/esferoide 3d.png}
%    \caption{DCE radiation from a spheroid rotating about an axis orthogonal to its symmetry axis. Emitted photon pairs satisfy %$\omega_1+\omega_2=2\Omega$.} 
%    \label{spheroid}
%\end{figure}
%{\color{red} Sugestões: - Adicionar setas laterais (diagonal esquerda-cima, direita-cima) representando um par de fotons com frequencias $\omega_1$ e $\omega_2$. Reduzir proporcionalmente o tamanho do spheroid, não precisa ser tão grande - Opcional(para caprichar ainda mais): aumentar o tamamnho do pequeno circulo indicando rotação, e aumentar ligeramente as fontes de $\Omega$, $r_\perp$ e $r_{/\!/}$
%. The radii parallel and perpendicular to the symmetry axis are denoted $r_\parallel$ and $r_\perp,$ respectively.
%Prolate spheroid rotating with angular frequency $\Omega$ about an axis orthogonal to its symmetry axis.}

\subsection{Anisotropy function and the case of SiO$_2$}

Using the analytical polarizability of a dielectric spheroid at rest~\cite{landau1946electrodynamics}, we obtain the anisotropy function $\Delta(\omega)$ governing the rotational DCE. For a prolate spheroid with radii $r_\perp$ and $r_\parallel$ ($r_\parallel>r_\perp$), we find 
\begin{equation}
    \frac{\Delta(\omega)}{4\pi\epsilon_0} = \frac{ r_\parallel r^2_\perp(N^{-1}_\parallel-N^{-1}_\perp)(\epsilon(\omega) - 1)^2}{6(\epsilon(\omega) - 1 + N^{-1}_\parallel)(\epsilon(\omega) - 1 + N^{-1}_\perp)},
    \label{deltadi}
\end{equation}
where $\epsilon(\om)$ is the dielectric function of the material and $N_i$ are the depolarizing factors, defined as functions of the eccentricity $e = \sqrt{1 - r^2_\perp/r^2_\parallel}$:

\begin{equation}
    \begin{split}\label{depo}
        N_\parallel &= \frac{1 - e^2}{2e^3}\left[\ln\left(\frac{1 + e}{1 - e}\right) - 2e\right], \\
        N_\perp &= \frac{1}{2}(1 - N_\parallel).
    \end{split}
\end{equation}

Because the lowest resonance frequency of SiO$_2$  lies in the THz range, 
its dielectric function can be approximated by $\epsilon(0)$ when considering \eqref{Gamma} for
the GHz rotation frequencies achieved in recent optical levitation experiments~\cite{Ahn2018,Reimann2018,Ahn2020,Jin2021,Zielinska2024}. The emission rate is then obtained from the quasi-static limit~\eqref{gammaqs} in terms of the anisotropy function at zero frequency.

As an example, we consider the parameters of the experiment reported in Ref.~\cite{Ahn2020}, with 
 $\Omega/(2\pi)=5.2\,{\rm GHz}$. 
As a crude approximation, we model the SiO$_2$ nanodumbbell as a prolate spheroid with semi-axes
$r_{\parallel}=2r_{\perp}= D,$ where $D= 150\,{\rm nm}$ is the diameter of the two nanospheres that are connected as a composite. 
By computing $\Delta(0)$ from \eqref{deltadi}-\eqref{depo} with $\epsilon(0)=3.9$ and plugging the result into \eqref{gammaqs}, we  obtain the extremely small emission rate $\Gamma_{\rm qs}= \scinot{2.5}{-21}\,{\rm s}^{-1}$.  Next, 
we show how the emission rate is enhanced by operating beyond the quasi-static regime exploiting a GHz polaritonic resonance.

 \subsection{Polaritonic-enhanced DCE from a BST nanoparticle}

 Materials with polaritonic resonances in the GHz range can strongly enhance the emission rate. BST provides a natural example, because its resonance frequency lies close to the highest experimentally achieved rotation frequencies~\cite{Ahn2020,Jin2021,Xu2021}.  
In this subsection, we consider a BST spheroid with the same dimensions discussed in the previous paragraph. 

We model the dielectric response of BST as an isotropic material using a single-resonance Lorentz function
\be\label{epsilon}
\epsilon(\om)=\epsilon_{\rm UV}+\frac{\epsilon(0)-\epsilon_{\rm UV}}{1-(\om/\om_T)^2+i\om\gamma/\om_T^2}
\ee
 with experimentally-measured parameters taken from Refs.~\cite{Turky2015,Xu2021}: resonance frequency $\om_T=\scinot{5.7}{9}[rad/s]$, damping rate $\gamma = \scinot{2.8}{8}[Hz]$, static permittivity $\epsilon(0)=7.1$, and high-frequency permittivity $\epsilon_{\rm UV}=2.896$. This single-oscillator model [Eq.~\eqref{epsilon}] is sufficiently accurate for our purposes, because we only consider frequencies $\omega$ in the GHz range, well below additional resonances that typically lie in the UV domain. The net effect of these UV resonances is nevertheless accounted for by the constant term $\epsilon_{\rm UV}$ in Eq.~\eqref{epsilon}, ensuring consistency with the expected static and high-frequency permittivities.

The quasi-static limit,  corresponding to very slow rotations with
$\Omega\ll \omega_T,$ offers a valuable benchmark for the full theory.
From Eq.~\eqref{gammaqs}, we obtain 
\begin{equation}\label{eq_qs_BST}
   \Gamma_{\rm qs} = 2.0\times 10^{-44}\,(\Omega[\mbox{GHz}])^7\,{\rm s}^{-1}.
\end{equation}

\begin{figure}[ht]
    %\centering
    \includegraphics[width=\columnwidth]{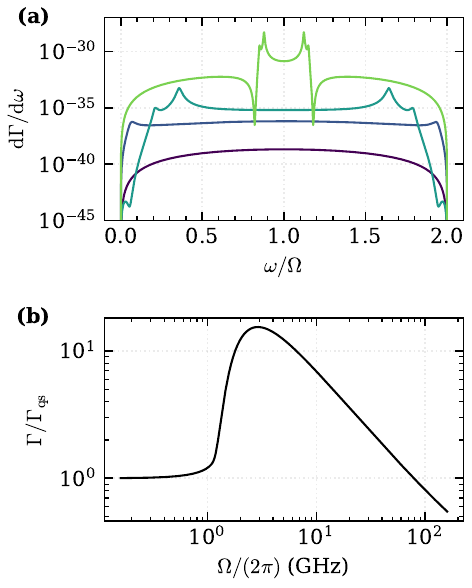}%/SpinningFiguraPrint_FI.pdf}
    \caption{ DCE emission of a prolate BST spheroid rotating about an axis orthogonal to its symmetry axis. (a) Emission spectrum for 
    (from top to bottom) 
  $\Omega/\omega_T=10,1.9,1.3,0.5$, where $\omega_T/(2\pi)= {0.91}\,{\rm GHz}$ is the BST resonance frequency. (b) Enhancement ratio (with respect to the quasi-static limit) $\Gamma/\Gamma_{\rm qs}$ versus rotational frequency $\Omega/(2\pi)$. }
    \label{fig:espectro_emissao}
\end{figure}

 We compute the emission spectrum by combining Eqs.~\eqref{spectrumEq}, \eqref{deltadi}, and \eqref{epsilon}. Figure~\ref{fig:espectro_emissao}(a) shows the spectrum for several values of $\Omega/\omega_T$ and reveals a structural change as $\Omega$ exceeds the resonance frequency $\omega_T$. For $\Omega=0.5\,\omega_T$, the spectrum remains close to the smooth polynomial form of the quasi-static limit. As $\Omega$ approaches $\omega_T$, resonance peaks inherited from the anisotropy function $\Delta(\omega)$ emerge. These peaks become progressively narrower as $\Omega$ increases further.

 Using this model, we compare the exact spectrum-integrated result for the emission rate $\Gamma$~\eqref{Gamma} with the quasi-static limit [Eqs.~\eqref{gammaqs} and \eqref{eq_qs_BST}]. Figure~\ref{fig:espectro_emissao}(b) shows the enhancement factor $\Gamma/\Gamma_{\rm qs}$ versus angular velocity $\Omega$. The resonant enhancement is clearly identified by a peak at $\Om \sim \om_T$. As the angular velocity increases beyond resonance, the enhancement $\Gamma/\Gamma_{\rm qs}$ decays as $1/\Omega$, because in the high-frequency regime ($\Omega\gg\omega_T$) the emission rate scales as $\Gamma\sim\Omega^6$.

\subsection{Optimal Geometry and Fundamental Limits under a Material Burst-Speed Constraint}

 We now optimize the spheroid size and shape to maximize DCE radiation, treating the spheroid eccentricity, radius and spinning frequency as free parameters. A key constraint is that the linear circumferential velocity of a spinning solid particle is bounded 
by the burst speed $v_b$ characteristic of a given material~\cite{Schuck2018}. 
In a spinning prolate spheroid, the fastest point is at a distance $r_\parallel$ from the rotation axis, and therefore the angular velocity is restricted by
$\Omega\, r_\parallel\le v_b=\sqrt{\frac{\rm UTS}{\rho}},$ where ${\rm UTS}$ is the  ultimate tensile strength and $\rho$ is the density. 

In order to estimate an upper bound for the emission rate, we 
consider the ${\rm UTS}$ of carbon nanotubes, which is approximately 
$\sim 10^3$ higher than in the case of SiO$_2.$
The burst speed is then $\sim 30$ higher than the value achieved in Ref.~\cite{Ahn2020} for SiO$_2$ nanoparticles, which leads to
the upper bound estimation of $v_b=1.5\times 10^5\,{\rm m/s}.$ 

\begin{figure}[ht]
    \centering
    \includegraphics[width=1\linewidth]{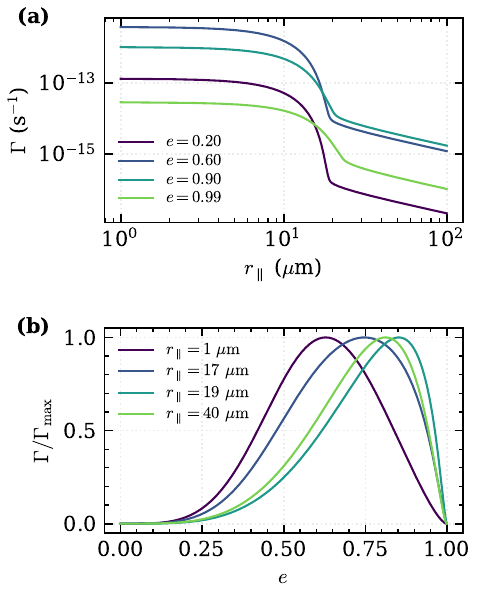}
    \caption{DCE emission from a spheroid at fixed tip velocity $v_b=\Omega r_{\parallel}=1.5\times10^5\,\mathrm{m/s}$. (a) Emission rate as a function of the major semi-axis $r_{\parallel}$ for several eccentricities $e$. 
    The crossover from the small-size limit (high frequencies) to the large-size one (low frequencies) takes place at $r_{\parallel}\sim v_b/\omega_T=26\,\mu{\rm m}.$
    The range shown spans rotational frequencies $\Omega/(2\pi)$ from $24\,\mathrm{GHz}$ to $0.24\,\mathrm{GHz}$. (b) Emission rate as a function of the eccentricity $e$ for different values of $r_{\parallel}.$ Each curve is normalized by its maximum value.}
    \label{fig:optimized}
\end{figure}

We consider the angular velocity at its limit value $\Omega =v_b/r_{\parallel}$  and plot the resulting emission rate $\Gamma$ as a function of $r_{\parallel}$ for different values of eccentricity in Fig.~\ref{fig:optimized}(a).
The dielectric function of the material is the same as considered in the previous sub-section.

From \eqref{Gamma}   and  \eqref{deltadi},
$\Gamma$ scales as $r_{\parallel}^6$
at a  fixed eccentricity. The nontrivial dependence 
shown in Fig.~\ref{fig:optimized}(b) arises from the constraint $\Omega=v_b/r_{\parallel}.$ The plateau for small particles shown in 
Fig.~\ref{fig:optimized}(a) corresponds to the high-frequency limit $\Omega\gg \omega_T$ for which 
$\Gamma \sim \Omega^6$ thus cancelling the 
dependence on $r_{\parallel}$ for small particles.

The crossover from the high-frequency limit to the quasi-static regime takes place 
at $r_{\parallel}\sim 10\,\mu{\rm m}.$
For larger particles, the frequencies are such that 
$\omega\ll \omega_T$ and then from 
\eqref{gammaqs}
$\Gamma\sim r_{\parallel}^6\Omega^7 \sim 1/r_{\parallel}$ as illustrated by Fig.~\ref{fig:optimized}(a).
The overall conclusion is that small particles such that $r_{\parallel}<v_b/\omega_T$ lead to a higher emission rate (for a given tip linear velocity $v_b$). However, using smaller particles require 
driving the rotation at a higher frequency in order to reach the linear velocity limit underlying our calculation. 

The comparison between different curves in Fig.~\ref{fig:optimized}(a) shows that more  ellongated particles tend to optimize the emission rate in the case of large particles, except for the extreme `needle-like' case $e=0.99.$ On the other hand,   the intermediate value $e=0.6$ leads to the highest rate
for small particles,
among the examples shown in the figure.

We investigate in more detail how the particle's shape can be optimized in Fig.~\ref{fig:optimized}(b), which shows the variation
of the emission rate  with the eccentricity for different values of $r_{\parallel}.$ As the size increases, the optimal value is displaced towards more elongated spheroids, but when reaching the quasi-static limit [$r_{\parallel} > 20\,\mu{\rm m}$ according to Fig.~\ref{fig:optimized}(a)], 
such trend is slightly reversed.

%%%%%%%%%%%%%%%%%%%%%%%%%%%%%%%%%%%%%%%%%%
\section{Conclusions}

In this work, we have shown that a rapidly rotating non-spherical dielectric nanoparticle can emit photons through the dynamical Casimir effect. We developed a theory based on the scattering of the full electromagnetic field by the spinning particle, where rotation generates frequency sidebands and induces Bogoliubov mixing of field modes. The emission is governed by the spectral anisotropy of the polarizability response, which depends critically on both particle geometry and material properties.

We investigated the corresponding DCE emission in a typical setting of optically levitated SiO$_2$ nanoparticles spinning at GHz frequencies. This corresponds to the quasi-static regime, where emission is determined by the static anisotropy and the rotation frequency. The presence of polaritonic resonances near the rotation frequency can lift this limitation and significantly enhance the emission rate. We provide realistic estimates for BST nanospheroids, which exhibit such resonances in the GHz range.

Finally, we show that the attainable emission is fundamentally constrained by the material burst velocity, which sets a maximum angular frequency for a given nanoparticle geometry. By optimizing the geometry of BST nanospheroids under this constraint, we determine the corresponding upper bounds on the DCE emission.

Despite these enhancements, the predicted rates remain exceedingly small, thereby establishing stringent quantitative limits on free-space rotational dynamical Casimir emission with a single particle. These results indicate that, in the absence of alternative amplification mechanisms or different physical platforms, direct observation of this effect in free space is unlikely with current or foreseeable experimental capabilities.

\section*{Acknowledgments}

We thank Helena Amaral,  Thiago Guerreiro, Laura Stolze, and Joanna Zieli\'nska for discussions. 
 This work was partially supported by Conselho Nacional de Desenvolvimento Cient\'{\i}fico e Tecnol\'ogico (CNPq--Brazil), Coordenaç\~ao de Aperfeiçamento de Pessoal de N\'{\i}vel Superior (CAPES--Brazil) and 
 Fundação Carlos Chagas Filho de Amparo à Pesquisa do Estado do Rio de Janeiro (FAPERJ--Brazil). JNM acknowledges partial support from a Limitless Space Institute Interstellar Initiative Grant.

\appendix

 \section{Polarizability tensor of a rotating anisotropic particle}

This Appendix serves two purposes. First, we provide the details of the derivation leading to the dipole response in Eq.~\eqref{drot}. Second, it provides an original expression for the polarizability of a spinning anisotropic particle. To our knowledge, previous results have been limited to rotation about a symmetry axis or to isotropic particles~\cite{Manjavacas2010}. 

We consider here an axisymmetric particle rotating about an axis orthogonal to its symmetry direction. This configuration is particularly relevant in view of recent experimental advances~\cite{Ahn2020}, where anisotropic dumbbell-like particles have reached record angular frequencies when rotated about an axis perpendicular to their symmetry axis.  The present derivation can be extended straightforwardly to more general anisotropic particles without symmetry.

As in the main text, we take a particle with a symmetry along the $\hat{\mathbf{x}}$ direction and rotating about the $\hat{\mathbf{z}}$ axis.  $\mathcal{R}$ denotes the laboratory frame and $\mathcal{R}'$ the co-rotating frame where the particle is at rest. In $\mathcal{R}'$, the dipole response is determined by the standard material response $\boldsymbol{\alpha}_{\rm rest}(\omega)$~[Eq.~\eqref{alpharest}] as
\begin{equation}\label{ind-rest}
\vb d'(\omega) = \boldsymbol{\alpha}_{\rm rest}(\omega)\cdot \vb E'(\omega),
\end{equation}
%where $\boldsymbol{\alpha}_{\rm rest}(\omega)$ is the standard material response defined in Eq.~\eqref{alpharest}.

The transformation from $\mathcal{R}'$  to $\mathcal{R}$ corresponds to the rotation $\vb r = \boldsymbol{\mathcal{R}}_z(t)\vb r'$, with
\begin{equation}
\boldsymbol{\mathcal{R}}_z(t)=
\begin{pmatrix}
\cos \Omega t & -\sin\Omega t & 0\\
\sin\Omega t & \cos \Omega t & 0\\
0 & 0 & 1
\end{pmatrix},
\end{equation}

Applying this transformation to the electric field relates the lab/co-rotating frame components in the Fourier domain
\begin{equation}
E_x(\omega)\pm iE_y(\omega)
=
E'_x(\omega\pm\Omega)\pm iE'_y(\omega\pm\Omega),
\label{Eiref}
\end{equation}
while the $z$ component remains unchanged. Equation~\eqref{Eiref} shows that rotation mixes field components at frequencies shifted by $\pm\Omega$.

Combining \eqref{Eiref} with the dipole relation in the co-rotating frame \eqref{ind-rest} and its shifted-frequency counterparts, we obtain \eqref{drot} for the dipole moment in the laboratory frame
 with
\begin{equation}
\boldsymbol{\alpha}_0(\omega)=
\begin{pmatrix}
\frac{1}{2}(S^+ + S^-) & \frac{i}{2}(S^+ - S^-) & 0\\
-\frac{i}{2}(S^+ - S^-) & \frac{1}{2}(S^+ + S^-) & 0\\
0 & 0 & \alpha_\perp(\omega)
\end{pmatrix}.
\end{equation}
We have introduced the polarizability 
\begin{equation}
S^{\pm}(\om)= \frac{\alpha_\parallel(\omega\pm\Omega)+\alpha_\perp(\omega\pm\Omega)}{2}
\label{sdef}
\end{equation}
 taken at the Doppler-shifted frequencies $\omega\pm\Omega$. Eq.~\eqref{drot} shows that the dipole response in the plane orthogonal to the rotation axis  at frequency $\omega$ is driven by the field modes at shifted frequencies $\omega\pm2\Omega$.

While our analysis focuses on a spheroidal nanoparticle to simplify the notation, the results extend to more general geometries. For an arbitrary dielectric particle, the polarizability tensor reads $\boldsymbol{\alpha}_{\rm rest}=\rm diag\,(\alpha_1,\alpha_2,\alpha_3)$ when taking the principal axis basis. If the particle rotates about a principal axis (e.g., axis 3 as in Fig.~\ref{spheroid}), Eqs. (\ref{drot})–(\ref{gammaqs}) still hold substituting $\alpha_1-\alpha_2$ for $\Delta=\alpha_{\parallel}-\alpha_{\perp}$. Thus, the frequency modulation at $2\Omega$ survives
even for asymmetric particles
as long as 
the rotation axis is not a symmetry axis
($\alpha_1 \neq \alpha_2$). This is a general feature within the dipole approximation and linear regime: because the polarizability $\alpha_{ij}$ is a rank-two tensor, it is inherently invariant under a $\pi$-rotation about any axis.

\end{document}